\documentclass[11pt]{article}
\usepackage{latexsym}
\usepackage{graphicx}
\usepackage{epstopdf}
	\DeclareGraphicsExtensions{.eps}
\usepackage{multicol}
\usepackage{caption}
\usepackage{amssymb}
\usepackage{amsmath}
\newtheorem{thm}{Theorem}[section]

\newtheorem{defn}{Definition}[section]

\newtheorem{exam}{Example}[section]

\newtheorem{rem}{Remark}[section]

\numberwithin{equation}{section}

\begin{document}

\vspace*{.5cm}
\begin{center}


{\LARGE{\textbf{ A short note on estimation of WCRE and WCE}}}
\bigskip

{\Large{\bf{Salimeh Yasaei Sekeh\footnote{Department\;of\;Statistics,\;Federal\;University\;of\;S$\tilde{\rm a}$o\;Carlos\;(UFSCar),\;S$\tilde{\rm a}$o\;Carlos,\;Brazil.\;E-mail: sa$_{-}$yasaei@yahoo.com}}}}




\end{center}



\mbox{\quad}
\def\fB{\mathfrak B}\def\fM{\mathfrak M}\def\fX{\mathfrak X}
 \def\cB{\mathcal B}\def\cM{\mathcal M}\def\cX{\mathcal X}
\def\bu{\mathbf u}\def\bv{\mathbf v}\def\bx{\mathbf x}\def\by{\mathbf y}
\def\om{\omega} \def\Om{\Omega}
\def\bbP{\mathbb P} \def\hw{h^{\rm w}} \def\hwphi{{h^{\rm w}_\phi}} \def\bbR{\mathbb R}
\def\beq{\begin{eqnarray}} \def\eeq{\end{eqnarray}}
\def\beqq{\begin{eqnarray*}} \def\eeqq{\end{eqnarray*}}
\def\rd{{\rm d}} \def\Dwphi{{D^{\rm w}_\phi}}
\def\BX{\mathbf{X}}
\def\mwe{{D^{\rm w}_\phi}}
\def\DwPhi{{D^{\rm w}_\Phi}} \def\iw{i^{\rm w}_{\phi}}
\def\bE{\mathbb{E}}
\def\1{{\mathbf 1}} \def\fB{{\mathfrak B}}  \def\fM{{\mathfrak M}}
\def\diy{\displaystyle} \def\bbE{{\mathbb E}} \def\bbP{{\mathbb P}}
\def\bF{\overline{F}}
\def\ew{{\mathcal{E}^{\rm w}_{\phi}}}
\def\cew{{\overline{\mathcal{E}}^{\rm w}_{\phi}}}
\def\lam{\lambda}
\let\phi\varphi

\begin{abstract}
In this note the author uses order statistics to estimate WCRE and WCE in terms of empirical and survival functions. An example in both cases normal and exponential WFs is analyzed.
\end{abstract}


\section{Introduction}
Let $X$ be a non-negative absolutely continuous random variable (RV) describing a component failure time, with the probability density function (PDF), $f(x)$, the cumulative distribution function (CDF), $F(x)=P(X\leq x)$, and the survival function (SF), $\bF(x)=P(X>x)$.

Following \cite{SY} for given  function $x\in \bbR\mapsto\phi (x )\geq 0$, and the PDF $f$,
the {\it weighted cumulative residual entropy} (WCRE) of $X$ (or $F$) with  weight function (WF) $\phi$ is defined by
\beq\label{eq:WCRE}
\ew (X)=\ew (F) =-\int_{\bbR^+}\phi (x )\bF(x)\log\,\bF(x)\rd x. \eeq
A standard agreement $0=0\cdot\log\,0=0\cdot\log\,\infty$ is adopted.
Given the CDF, $\bx\in\bbR^+ \mapsto F(\bx )\in [0,1]$, with WF $\phi$, the {\it weighted cumulative entropy} (WCE) of a RV $X$ is presented by
\beq\label{eq:WCE}
\cew (X)=\cew (F) =-\int_{\bbR^+}\phi (x )F(x) \log\,F(x)\rd x. \eeq
Note that in particular $\phi(x)=x$ the WCRE and WCE in (\ref{eq:WCRE}) and (\ref{eq:WCE}) turn (8) and (9) in \cite{M}. Evidentally for $\phi\equiv 1$, the WCRE and WCE take the forms CRE and CE, cf. \cite{CL}.

Passing to Kullback-Leibler divergence $D(.\|.)$ and CRE $\mathcal{E}(.)$, it's worthwhile to note that
\beq \ew(F)=\diy-D(\varphi \overline{F}\|\varphi)=\mathcal{E}(\varphi F)+D(\varphi \overline{F}\|\overline{F}).\eeq
\begin{rem}
{\rm (a)}\;Suppose an RV $X$ follows exponential distribution with mean $\diy\frac{1}{\lambda}$. Then
\beqq \ew(F)=\diy\frac{1}{\lambda}\bbE_Z\big[\varphi(Z)\big], \eeqq
where RV $Z$ has Gamma distribution with shape parameter 2 and scale $\diy\frac{1}{\lambda}$. Note that for any real number $\xi$ and particular form of $\varphi(x)=\diy e^{-2\pi i x\xi}$, the WCRE becomes the Fourier transform of Gamma's PDF. Moreover, if $\varphi(x)=\diy e^{itx}\Big(\diy e^{-s\;x}\Big),\; t\in\bbR$, then $\ew(F)$ takes the Gammas's characteristic function (Laplace transform). Also assume the case $\varphi$ is the polynomial function of $x$, $x\geq 0$, $a_i\geq 0,\;i=0\dots n$:
\beqq \varphi(x)=a_n\;x^n+a_{n-1}\;x^{n-1}+\dots+a_1\;x+a_0.\eeqq
Observe that the WCRE reads
\beqq \ew(F)=\diy\sum_{i=0}^n a_i \diy\frac{\Gamma(i+2)}{\lambda^{i+1}},\quad \Gamma(i+2)=(i+1)!. \eeqq
{\rm (b)}\;The weighted entropy defined in \cite{SY}, $h^{\rm w}_{\varphi}$, for the equilibrium RV, $X_e$, associated to a non-negative RV $X$ with SF $\bF$ relates to the WCRE by
\beqq \bbE[X]\;h^{\rm w}_{\varphi}(X_e)=\ew(F)+\log\bbE[X]\;\diy\int_{\bbR^+}\varphi(x)\bF(x)\rd x. \eeqq
\end{rem}
The reader would be addressed to \cite{SY} for numerous properties and motivations of WCRE and WCE. Also refer to \cite{BG}, \cite{SY1} for more details regarding the weighted entropies.

The main aim of this work is to represent an empirical estimator for WCRE and WCE, following analogue arguments in \cite{M}, \cite{RCVW}.

\section{Main results}
 The paper statistically focuses on the estimation of WCRE and WCE with general WF $\phi$ by using the order statistics. Here and below $\widehat{F}_n$ and $\widehat{\overline{F}}$ stand the empirical distribution and survival function of random sample $X_1,X_2,\dots,X_n$ at point $x$:
\beqq \widehat{F}_n(x)=\frac{1}{n} \sum\limits_{i=1}^n \1\{X_i\leq x\} \quad ,\quad \widehat{\overline{F}}_n(x)=1-\widehat{F}_n(x),\eeqq
where $\1\{X\leq x\}$ is the indicator function of the event $\{X\leq x\}$.
\begin{defn}
Let $X_1,X_2,\dots,X_n$ be a random sample drawn from a population with distribution function $F$. Again set $\phi(x)$ as WF, according to the definition of WCRE and WCE in (\ref{eq:WCRE}) and (\ref{eq:WCE}), the empirical WCRE and empirical WCE respectively are defined by
\beq \begin{array}{c}\label{eq:sec3.01}\diy \ew(\widehat{F}_n)=-\diy\int_0^\infty \phi(x) \widehat{\overline{F}}(x) \log \widehat{\overline{F}}(x) \rd x,\\
\diy \cew(\widehat{F}_n)=-\diy\int_0^\infty \phi(x) \widehat{F}(x) \log \widehat{F}(x) \rd x.\end{array}\eeq
\end{defn}
\vskip .5 truecm
Paying homage to the Theorem 9 in \cite{RCVW}, with similar methodology the following assertion holds true, omitting the proof.

\begin{thm}\label{thm1}
Let $X$ be a non-negative RV in $L^p$. Given $p>1$ suppose for given $0<a<\infty$ the WF $\phi$ obeys
\beq\label{assum:1} \int_0^a \phi(x)\rd x<\infty\;\; \textrm{and} \;\; \int_a^\infty \phi(x) x^{-p} \rd x <\infty.\eeq
Then the empirical WCRE convergens to the WCRE of $X$, that is :
\beqq \ew(\widehat{F}_n) \rightarrow \ew(F)\;\;\; a.s. \eeqq
\end{thm}
\vskip .5 truecm
Next, consider the order statistics of random sample, denote by $X_{(1)},X_{(2)},\dots,X_{(n)}$, indeed on the other hand $\cew(\widehat{F}_n)$ renders the empirical quantity as
\beq\label{eq:sec3.02} \cew(\widehat{F}_n)=-\diy \sum\limits_{i=1}^{n-1}\int_{X_{(i)}}^{X_{(i+1)}} \phi(x) \widehat{F}_n(x) \log \widehat{F}_n(x)\rd x.\eeq
Further, set
$$\zeta_i=i(\psi(x_{(i+1)})-\psi(x_{(i)})\quad \hbox{and}\quad \psi(x)=\diy \int_0^x \phi(t) \rd t.$$
Taking into account $\widehat{F}_n(x)=\diy\frac{i}{n}$, $X_{(i)}\leq x \leq X_{(i+1)}$, $i=1,\dots,n-1$, the Eqn (\ref{eq:sec3.02}) becomes
\beqq \begin{array}{ccl}
\cew(\widehat{F}_n)&=&\diy-\sum\limits_{i=1}^{n-1}\bigg(\psi(x_{(i+1)})-\psi(x_{(i)})\bigg) \frac{i}{n} \log \frac{i}{n}\\
&=&-\diy\frac{1}{n} \sum\limits_{i=1}^{n-1} \zeta_i [\log i-\log n].\end{array}\eeqq
In addition following more straightforward computation yields
\beqq\begin{array}{ccl}
\sum\limits_{i=1}^{n-1}\zeta_i&=&\diy \sum\limits_{i=1}^{n-1} \sum\limits_{k=1}^{i} \bigg(\psi(x_{(i+1)})-\psi(x_{(i)}\bigg)\\
&=&\diy \sum\limits_{k=1}^{n-1} \sum\limits_{i=k}^{n-1}\bigg(\psi(x_{(i+1)})-\psi(x_{(i)}\bigg)\\
&=&\diy \sum\limits_{k=1}^{n-1} \bigg(\psi(x_{(n)})-\psi(x_{(k)}\bigg)= n\bigg(\psi(x_{(n)})-\overline{\psi}\bigg),\end{array}\eeqq
where $\overline{\psi}=\frac{1}{n}\sum\limits_{i=1}^n \psi(x_{(i)})$. Consequently,
\beq\label{eq:sec3.03} \cew(\widehat{F}_n)=\big(\psi(x_{(n)})-\overline{\psi}\big)\log n-\frac{1}{n} \sum\limits_{i=1}^{n-1} \zeta_i\log i.\eeq
Now set $\tau_i=(n-i)\big(\psi(x_{(i+1)})-\psi(x_{(i)}\big)$,  similarly it is plain that
\beq\label{eq:sec3.04} \ew(\widehat{F}_n)=\big(\overline{\psi}-\psi(x_{(1)})\big)\log n-\frac{1}{n} \sum\limits_{i=1}^{n-1} \tau_i \log(n-i).\eeq

The note is concluded by giving an example in order to discuss the represented estimation for WCRE and WCE above. In fact in Example \ref{exam.Simulation}, two Normal and Exponential WFs are considered. The observations from the simulations are coincide with analytical computation here and in \cite{SY}.
\begin{figure}[!htbp]
\begin{center}
\includegraphics[scale=1]{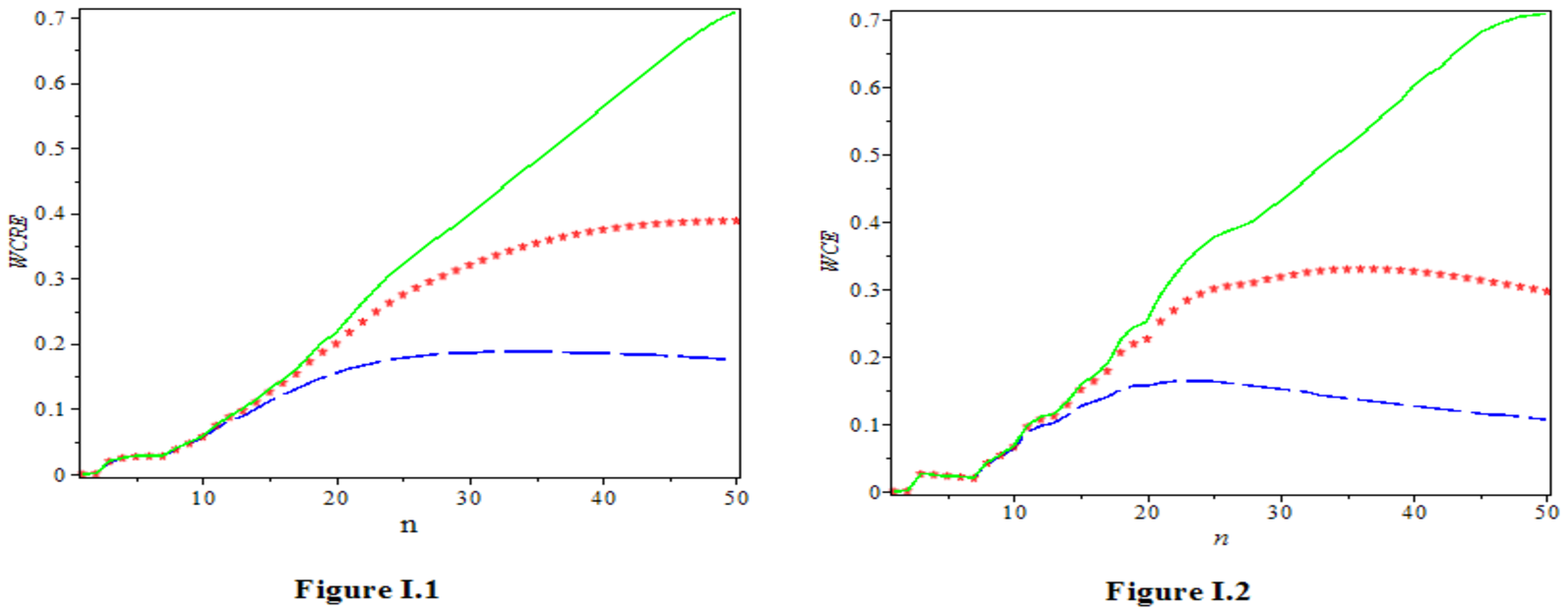}
\end{center}
\caption*{{\small Empirical WCRE and WCE for the data of Example \ref{exam.Simulation} with Normal WF: Longdash ($\sigma=0.5$), point ($\sigma=1$) and Solid ($\sigma=2$).}}
\label{fig}
\end{figure}
\begin{exam}\label{exam.Simulation}
{\rm For Given set of 50 Sample lifetime generated from exponential distributions with mean 2:\\
\beqq\begin{array}{cccccccccc}
  8.23 & 2.86 & 0.906 & 6.66 & 0.912 & 0.127 & 0.290 & 0.422 & 7.23 & 10.93 \\
  0.126 & 1.65 & 2.18 & 1.18 & 1.43 & 0.521 & 1.34 & 0.428 & 3.40 & 3.36 \\
  0.119 & 1.83 & 1.24 & 1.37 & 4.54 & 6.44 & 0.626 & 2.37 & 0.906 & 1.72 \\
  0.049 & 1.51 & 0.123 & 0.651 & 3.32 & 1.42 & 2.74 & 1.96 & 0.047 & 0.120 \\
  0.247 & 0.417 & 4.24 & 2.04 & 6.01 & 0.721 & 2.57 & 1.45 & 2.22 & 0.221
\end{array}\eeqq

{\bf (I)\,Normal WF:}\; According to the upcoming sample, assume the WF
$$\phi(x)=\diy\exp\big(-{x^2}\big/{2\sigma^2}\big).$$
The graphs in Figure I.1, I.2 present the empirical WCRE and WCE for the given data in particular value of $\sigma=0.5,\;1,\;2$. These simulations show that both WCRE and WCE are increasing in $\sigma$, however this is what we were expecting from the analytical calculations. In addition, for biggest $\sigma=2$, same as CRE and CE, when the number of sample order data, $n$, is increasing the both curves WCRE and WCE are ascending as well, whereas for small value of $\sigma$ this is violated. For instant for $\sigma=0.5$, $WCRE(35)=0.1872$ and $WCRE(40)=0.1847$ or when $\sigma=1$, $WCE(35)=0.3299$ and $WCE(40)=0.3270$.\\
Here $WCRE(n)$ and $WCE(n)$ are the amount of WCRE and WCE in $n$ respectively.}\\

{\rm {\bf (II)\,Exponential WF:}\; Suppose that $\phi(x)=\exp(tx)$. To satisfy the condition (\ref{assum:1}) in Theorem \ref{thm1}, constant $t$ should be considered non-positive. Observe the Figure II.1, II.2 for the given sample data. The graphs represents the WCRE and WCE for different values of $t=-1,\;-0.2,\;-0.0001$. Here also the WCRE and WCE are increasing when $t$ is increasing.
Although in spite of Normal WF case, for each value of $t$ the WCRE and WCE increase singly.}
\end{exam}
\begin{figure}[t!]
\begin{center}
\includegraphics[scale=1]{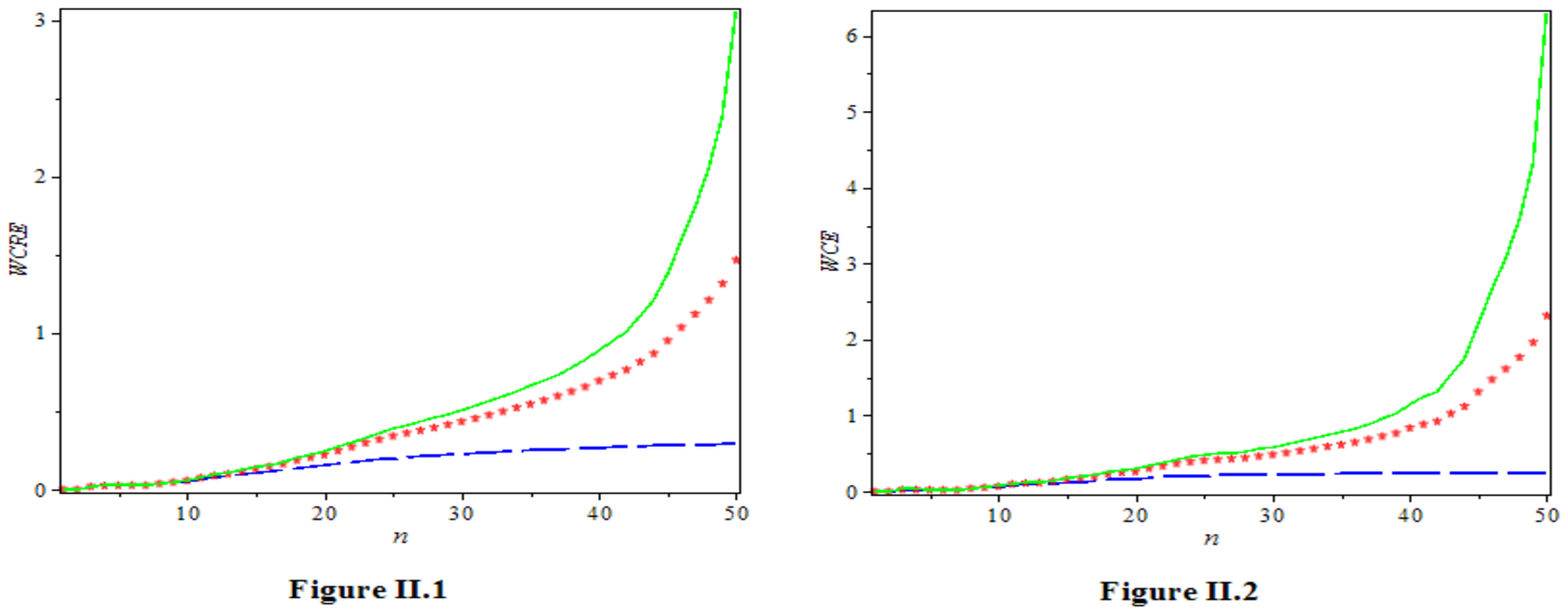}
\end{center}
\caption*{{\small Empirical WCRE and WCE for the data of Example \ref{exam.Simulation} with exponential WF: Longdash ($t=-1$), point ($t=-0.2$) and Solid ($t=-0.0001$).}}
\label{fig}
\end{figure}
\vskip .5 truecm
{\emph{Acknowledgements --}}
SYS thanks Professor Yuri Suhov for useful discussions. SYS also thanks the CAPES PNPD-UFSCAR Foundation
for the financial support in the year 2014-5 and the Federal University of Sao Carlos, Department of Statistics, for hospitality during the year 2014-5.


\end{document}